\documentclass[journal,12pt,onecolumn,draftclsnofoot,]{IEEEtran}
\ifCLASSINFOpdf
\else
\fi

\usepackage{dblfloatfix}

\usepackage{graphicx}
\usepackage{textcomp}
\usepackage{lipsum}

\usepackage{lipsum}
\usepackage{graphicx}

\usepackage{subfig}
\usepackage{graphicx}
\usepackage{caption}

\usepackage{balance}

\usepackage{amsmath}

\usepackage{multirow}

\usepackage{cite}

\usepackage{csquotes}

\usepackage{mathtools}

\usepackage{amsmath}

\DeclareMathOperator{\erfc}{erfc}
\DeclareMathOperator{\SNR}{SNR}
\DeclareMathOperator{\m}{m}

\DeclareMathOperator{\pd}{pd}
\DeclareMathOperator{\eff}{eff}
\DeclareMathOperator{\shot}{shot}
\DeclareMathOperator{\thermal}{thermal}
\DeclareMathOperator{\A}{A}
\DeclareMathOperator{\peak}{peak}
\DeclareMathOperator{\cm}{cm}
\DeclareMathOperator{\Hz}{Hz}
\DeclareMathOperator{\J}{J}
\DeclareMathOperator{\K}{K}
\DeclareMathOperator{\dB}{dB}

\usepackage[table]{xcolor}
\definecolor{coolblack}{rgb}{0.0, 0.18, 0.39}
\definecolor{applegreen}{rgb}{0.55, 0.71, 0.0}
\definecolor{mediumcarmine}{rgb}{0.69, 0.25, 0.21}
\definecolor{mediumlavendermagenta}{rgb}{0.8, 0.6, 0.8}
\definecolor{selectiveyellow}{rgb}{1.0, 0.73, 0.0}
\definecolor{blue}{rgb}{0,0,1}

\begin{document}
%
\title{Inter-Satellite Communication System based on Visible Light}
%
%
%

\author{David~N.~Amanor,~\IEEEmembership{Member,~IEEE,}
        William~W.~Edmonson,~\IEEEmembership{Senior Member,~IEEE,}
        and~Fatemeh~Afghah,~\IEEEmembership{Member,~IEEE}
\thanks{D. N. Amanor was with the Department of Electrical and Computer Engineering, North Carolina A\&T State University, Greensboro, NC 27411, USA. He is now with Intel Corporation, Hillsboro, OR 97124, USA.
 e-mail: narh.amanor@gmail.com}
\thanks{W. W. Edmonson is a National Institute of Aerospace (NIA) Langley Professor, and is with the Department
of Electrical and Computer Engineering, North Carolina A\&T State University, Greensboro,
NC 27411, USA. e-mail: wwedmons@ncat.edu}
\thanks{F. Afghah is with the School of Informatics, Computing \& Cyber Systems, Northern Arizona University, Flagstaff, Arizona 86011,
USA. e-mail: fatemeh.afghah@nau.edu}
\thanks{Manuscript received August 21, 2017; revised February 12, 2018.}}

%
%

\markboth{IEEE Transactions on Aerospace and Electronic Systems,~Vol.~x, No.~x, xxxxx~2018}%
{Shell \MakeLowercase{\textit{et al.}}: Bare Demo of IEEEtran.cls for IEEE Journals}
%



\maketitle


\begin{abstract}

Future space missions will be driven by factors such as the need for reduced cost of spacecraft without diminished performance, new services and capabilities including reconfigurability, autonomous operations, target observation with improved resolution and servicing (or proximity) operations. Small satellites, deployed as a sensor network in space, can through inter-satellite communication (ISC) enable the realization of these future goals. Developing the communication subsystem that can facilitate ISC within this distributed network of small satellites require a complex range of design trade-offs. For small satellites, the general design parameters that are to be optimized for ISC are size, mass, and power, as well as cost (SMaP-C). Novel and efficient design techniques for implementing the communication subsystem are crucial for building multiple small satellite networks with capability for achieving significant data-rates along the inter-satellite links (ISLs). In this paper, we propose an alternative approach to RF and laser ISLs for ISC among small satellites deployed as a sensor network in low Earth orbit (LEO). For short to medium range ISLs, we present an LED-based visible light communication (VLC) system that addresses the SMaP constraints, including capability for achieving significant data rates. Our research is focused on the development of the physical layer for pico-/nano class of satellites with prime consideration for the impact of solar background illumination on link performance. We develop an analytical model of the inter-satellite link (ISL) in MATLAB and evaluate its feasibility and performance for different intensity modulation and direct detection (IM/DD) schemes. Using a transmitted optical power of 4W and digital pulse interval modulation (DPIM), a receiver bandwidth requirement of 3.5 MHz is needed to achieve a data rate of 2.0 Mbits/s over a moderate link distance of 0.5 km at a BER of 10\textsuperscript{-6}.

\end{abstract}


%
\IEEEpeerreviewmaketitle

\section{INTRODUCTION}

\IEEEPARstart{T}{he} development of small-size, light-weight, low-power and low-cost satellites has witnessed significant growth in the last few years. An important class of small satellites, which is being used by academia, industry and government as a platform for space exploration and research, is CubeSats. These satellites  are special category of nanosatellites defined in terms of 10 cm $ \times $ 10 cm $ \times $ 10 cm  sized units (approx. 1.3 kg each) called ``U\textquotesingle s''. Although a 1U CubeSat can be extended to higher configuration (i.e., 1.5, 2, 3, 6, and 12U) if more capability is required, it is crucial to resist the creep toward larger and more expensive CubeSat missions, as this defeats the primary goal of maintaining low-cost approaches as the cornerstone of CubeSat development \cite{academy}. 

Small satellites, deployed as a sensor network in space, have an advantage over conventional satellites in space exploration because of their potential to perform coordinated observations, high-resolution measurements, and identification of Earth\textquotesingle s asset that is inclusive of its space environment. Low-latency communications between these satellites result in improved availability for observation, telecommunications and reconnaissance applications \cite{conney}. One fundamental reason for the shift from using large and expensive satellites to multiple low-cost small satellites is the resulting inherent intelligence of the distributed multi-satellite nodes which has potential for autonomous operations. Another driving motivation for the development of large constellations of small satellites is the desire for rapid revisit rates or persistence from low-earth-orbit (LEO) satellites. Such satellites have the potential to provide inter-satellite data relay, providing a highly survivable mesh of nodes capable of relaying data before downlink to ground stations \cite{conney}. To enable cooperation among these distributed multi-satellite nodes requires a need for inter-satellite communication. Presently, the dominant research and development for implementing inter-satellite communication links (ISLs) consists of using either radio frequency (RF) or highly directed lasers. The latter will require a highly accurate pointing satellite control system, while the former is not suitable for systems with sensitive electronics onboard or in applications where high data rates are required due to the limited available spectrum. It is also mechanically challenging to deploy large parabolic antennas on small satellites equipped with RF radios in order to support high data rates. The required pointing accuracy needed for laser communication presents a challenge to the form factor of pico-/nano class of satellites due to the stringent SMaP restrictions imposed by the platform. Lasers produce a narrow and focused beam of light that could fall out of the field-of-view (FOV) of a small satellite receiver due to slight movements. Furthermore, for formation flying systems in LEO, the ISLs are much shorter than links between satellites in geostationary orbit; thus, the use of lasers and the highly accurate pointing they provide can be considered superfluous\cite{wood},\cite{amanor}. To minimize the SMaP constraints imposed by the platform, along with the need for reduced pointing accuracy and to achieve high data transmission rates, we propose a visible light communication (VLC) subsystem for pico-/nano class of satellites for ISC. These multiple small satellite missions will benefit from VLC\textquotesingle s ability to transmit higher data rates with smaller, light-weight nodes, while avoiding the usual interference problems associated with RF, as well as the apparent radio spectrum scarcity below the 6 GHz band \cite{rajagopal}. Furthermore, the electronics required for achieving precision pointing accuracy for laser communication systems will be avoided. With approximately 300 THz of free bandwidth available for VLC, high capacity data transmission rates could be provided over short distances using arrays of LEDs.

\begin{figure}[!t]
\centering
	\includegraphics[width=3.5in]{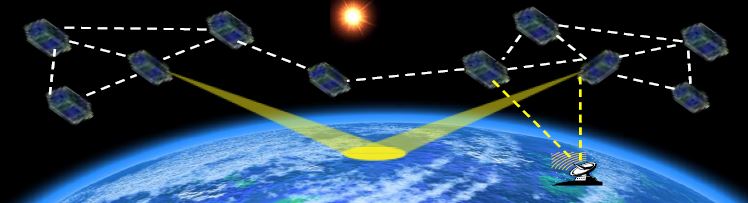}
	\caption{Network of Small Satellites: Adapted from \cite{amanor2}}
\label{fig 1}
\end{figure}

This paper is an extensive treatment of the preliminary study presented in \cite{amanor} and \cite{amanor4s}. In these conference papers, we proposed a high-level description of a VLC system for ISC among small satellites. In this paper, we developed the physical layer requirements and design concepts for a VLC-based communication subsystem for ISC in small satellite networks. The proposed system addresses the SMaP constraints of small satellites and challenges associated with RF and Laser ISLs in small satellites networks.

The remainder of the paper is structured as follows. Section II covered background information of related works on ISC. The design considerations and proposed system description are presented in Sections III and IV, respectively.  Section V examined the VLC link physical model, while Section VI treated the solar background noise model. The characteristics of the VLC modulated signal is discussed in Section VII, followed by an example power budget design in Section VIII. The performance evaluation of an analytical model of the proposed system is treated in Section IX, and the concluding remarks are presented in Section X.

\section{BACKGROUND}

Most of the launched and projected missions of multiple small satellite systems employed RF or laser ISLs \cite{radakrishnan2}. Among these missions, the most ambitious one is QB-50, which uses RF ISLs and consists of a network of CubeSats that will study the Earth\textquotesingle s upper thermosphere, measuring oxygen levels, and electron behavior among others. All 50 CubeSats were supposed to be launched together in February 2016, but due to the unavailability of the launch vehicle, the plan was revised and 28  CubeSats were deployed from the International Space Station (ISS) in May 2017, followed by the launch of another 8 CubeSats from an Indian Polar Satellite Launch Vehicle (PSLV) in late May 2017.

Notwithstanding the dominance of RF and laser ISLs in most multiple small satellite missions, recent advancements in LED technology have triggered renewed interest in VLC as a viable alternative to RF and laser for LOS communication links of moderate scope. Visible light communication systems exploit the optical bandwidth available within the visible light band (i.e., 380 nm to 750 nm) for data communications. LED-based transmitter sources have a relative advantage over RF and laser transmission sources due to their low power requirements, light-weight, and small footprint. In \cite{wood}, the feasibility of LEDs for short-range ISLs was examined for a hypothetical low-end ISL. The work discussed methods for minimizing background illumination, but did not quantitatively evaluate solar background illumination and its impact on ISL performance. The fundamental analysis for VLC system using LED lights for indoor applications was discussed in \cite{komine}. In \cite{nakajima}, a VLC system using LEDs was successfully demonstrated between satellite and ground. The ShindaiSat, Shinshu University Satellite, is a VLC experimental satellite for on-orbit technology demonstration using LED light as a communication link. To achieve this feat, the ShindaiSat used a relatively large micro-satellite measuring 400 mm $\times$ 400 mm $\times$ 450 mm and weighing 35 kg. This contrasts sharply with the form-factor of CubeSats. In \cite{arruego}, LEDs were evaluated and flown in orbit for intra-satellite communication between internal assemblies’ onboard satellites. These lamps combine very low-power consumption with an extremely long operational life, maintaining during all their operation the same chromaticity without significant changes.
 
The above studies and on-orbit demonstrations on LED-based VLC underscores the potential feasibility of this technology for ISC. However, investigating the feasibility of LEDs for ISC without a quantitative evaluation of solar background radiation that reaches the receiver field-of-view (FOV) and its impact on the SNR leaves a research gap that needs to be addressed. This is because the radiative energy that the Sun emits within the visible light spectrum (i.e., 380 nm - 750 nm) to the Earth system, including LEO, is about 595 W/$\m^2$. This high background illumination is large enough to ``drown" the received information signal from an LED source. 

By modeling the solar background power and numerically evaluating its impact on the ISL, this paper seeks to fill the research gap in previous studies, and demonstrate the feasibility of using LED-based visible light links for ISC in future multiple small satellite space missions.

\begin{figure*}[!t]
\centering
  \includegraphics[width=12cm,height=2.7cm]{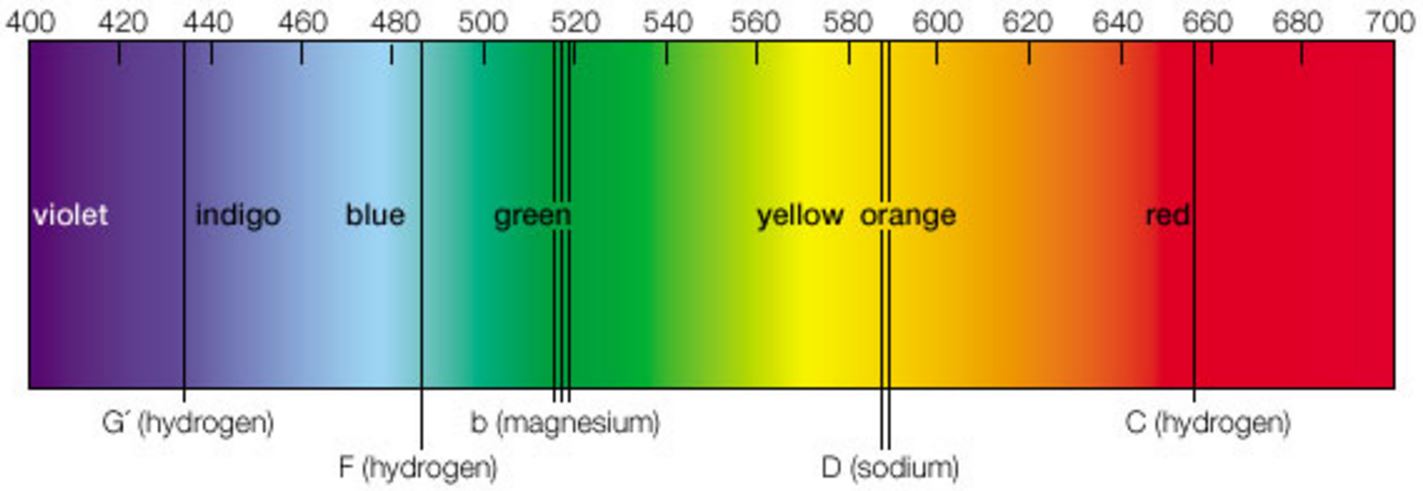}
  \caption{Fraunhofer Lines within Visible Light Spectrum \cite{britannica}: \textsl{wavelength} (nm)}
\label{fig2}
\end{figure*}

\section{KEY DESIGN CONSIDERATIONS}

In general, a satellite whether small or large, is composed of several functional subsystems including communications, attitude determination and control, tracking, telemetry and command (TTC), as well as electrical power supply. For small satellites, it is crucial for the subsystem\textquotesingle s designer to take into account the overall system\textquotesingle s SMaP constraints in order to avoid any violation of the stringent size and volume restrictions. The characteristics of the operational environment must also be considered. In this section, we summarized the critical design issues of LED-based VLC system for ISC among small satellites.

\subsection{APPROACH TO MITIGATE BACKGROUND RADIATION}

The radiative energy per cross-sectional unit area that the Sun emits to the Earth system across all wavelengths of the electromagnetic spectrum based on Planck\textquotesingle s radiation formula is 1360 W/$\m^2$. The equivalent radiative energy within the visible light band is approximately 595 W/$\m^2$. The severity of this background power is enough to degrade the SNR at the receiver and thus poses a threat to reliable visible light communications. However, at certain frequencies within the visible band, the Sun\textquotesingle s output spectrum has been absorbed by chemical elements present in the Sun, and in the process they leave a characteristic fingerprint on the solar spectrum in the form of black lines (i.e., Fraunhofer lines). The power and resulting noise from the Sun at these frequencies is reduced. By placing photodetectors behind optical filters selected to match Fraunhofer lines can enable clear signal detection even when the detector is directly facing the Sun \cite{wood}. At the most intense Fraunhofer lines, the solar background falls below 10 percent of its continuum values \cite{gelbwachs}.

This work is inspired by the background radiation mitigation concept espoused in \cite{wood} and the receiver design approach proposed in \cite{barry} to develop a noise-resistant inter-satellite communication system for small satellites using LED(s) at the transmitter and a photodetector at the receiver. The LED is chosen such that its peak transmission energy (or peak wavelength) lies at the center of a Fraunhofer line, while the optical front-end of the receiver consists of a filter whose passband matches the Fraunhofer line spectral width.  Some prominent Fraunhofer lines are illustrated in Fig. 2.

\subsection{DOPPLER EFFECTS}

A fundamental problem that needs to be addressed for ISC is Doppler effect and its impact on the ISLs. A Doppler shift causes the received signal frequency of a source to differ from the sent frequency due to motion that is increasing or decreasing the distance between the source and receiver. For our proposed application, the background signal from the Sun and the information carrying signal from the transmitting satellite will both experience some form of Doppler shifts at the receiver due to the continuous motion of the receiver that either increases or decreases the distance between the receiver and the two sources.  The impact of Doppler effects on the performance of inter-satellite links in LEO has been studied in \cite{liu2} and \cite{yang}. The normalized wavelength shift between a transmitting and receiving satellite is given by \cite{liu2}:


\begin{equation}\Delta \lambda= \frac{\lambda_s}{c} \frac {d}{dt}| r(t,\tau)| \end{equation}

where \begin{equation}\Delta \lambda= {\lambda_d - \lambda_s} \end{equation}
and \begin{math}\Delta \lambda \end{math} stands for normalized Doppler wavelength shift;  \begin{math}\lambda_d \end{math} and \begin{math}\lambda_s \end{math} are the wavelengths of the received signal and emitted signal, respectively. The term,\begin{math}\ r(t,\tau) \end{math}, represents the actual propagation range of the signal from the source satellite to destination.

It follows from (1) that a normalized Doppler shift of 0.015 nm corresponds to spacecraft moving at a relative velocity of 9 km/s for a 500 nm emitted light signal, while a shift of 0.05 nm corresponds to a relative velocity of 30 km/s. Shifts smaller than 0.001 nm are assumed to be insignificant \cite{kerr}.

In this work, we focus on intra-orbit ISLs, where the distance between satellites is fixed and Doppler effects is negligible.

\subsection{PROPOSED LED PEAK WAVELENGTHS FOR TRANSMISSION}

A limited number of Fraunhofer lines offer a natural low background noise channel for VLC. The wavelengths and bandwidths of the most intense Fraunhofer lines are shown in Table I.
We selected Fraunhofer lines with bandwidths greater than 250 GHz in order to ensure that Doppler shifts are accommodated within the Fraunhofer linewidth. The LED signal transmissions will be centered on these Fraunhofer lines and the bandwidths are broad enough to accommodate any Doppler shifts that may cause marginal shifts of the targeted Fraunhofer line without the need for additional on-board electronics to provide retuning.

The Fraunhofer lines in Table I possess significant bandwidth that can be exploited for high data rate ISLs. In particular, the Fraunhofer lines at 393.3682 nm and 396.8492 nm wavelengths are broad enough to guarantee stable transmissions even in the presence of sizable Doppler shifts. 

We did not consider Fraunhofer lines in the range 490 nm to 590 nm in the selection of potential frequencies (shown in Table I) for the proposed system due to their proximity to the Sun\textquotesingle s peak, which is close to the 500 nm wavelength mark. For Si PIN Photodiodes, transmissions along Fraunhofer lines below 390 nm wavelength may suffer from poor detector responsivity, and therefore would not be appropriate for applications where very weak signals reaches the detector. The proximity of these lines to the ultravoilet region also poses a hazard to terrestrial applications, but this may not be an issue for space applications.


\begin{table}[ht]
\caption{The Most Intense Solar Fraunhofer Lines with Bandwidth greater than 250 GHz \cite{lang, sethi}} 
\centering 
\begin{tabular}{c c c c} 
\hline\hline 
Wavelength\ & Spectral Width\ & Bandwidth\ & Element\ \\
nm & nm & GHz \\[0.5ex] 
\hline 
381.5851 & 0.1272 & 262.1 & Fe \\ 
382.0436 & 0.1712 & 351.9 & Fe \\
382.5891 & 0.1519 & 311.3 & Fe \\
383.2310 & 0.1685 & 344.2 & Mg \\
383.8302 & 0.1920 & 391.0 & Mg \\
385.9922 & 0.1554 & 312.9 & Fe \\
\cellcolor{gray!25}393.3682 & \cellcolor{gray!25}2.0253 & \cellcolor{gray!25}3926.6 & \cellcolor{gray!25}Ca \\
\cellcolor{gray!25}396.8492 & \cellcolor{gray!25}1.5467 & \cellcolor{gray!25}2946.3 & \cellcolor{gray!25}Ca \\
410.1748 & 0.3133 & 558.7& H \\
434.0475 & 0.2855 & 454.6& H \\
486.1342 & 0.3680 & 467.2& H \\
656.2808 & 0.4020 & 280.0& H \\ [1ex] 
\hline 
\end{tabular}
\label{table:table1} 
\end{table}


\begin{figure*}[!b]
\centering
  \includegraphics[width=15cm,height=5.0cm]{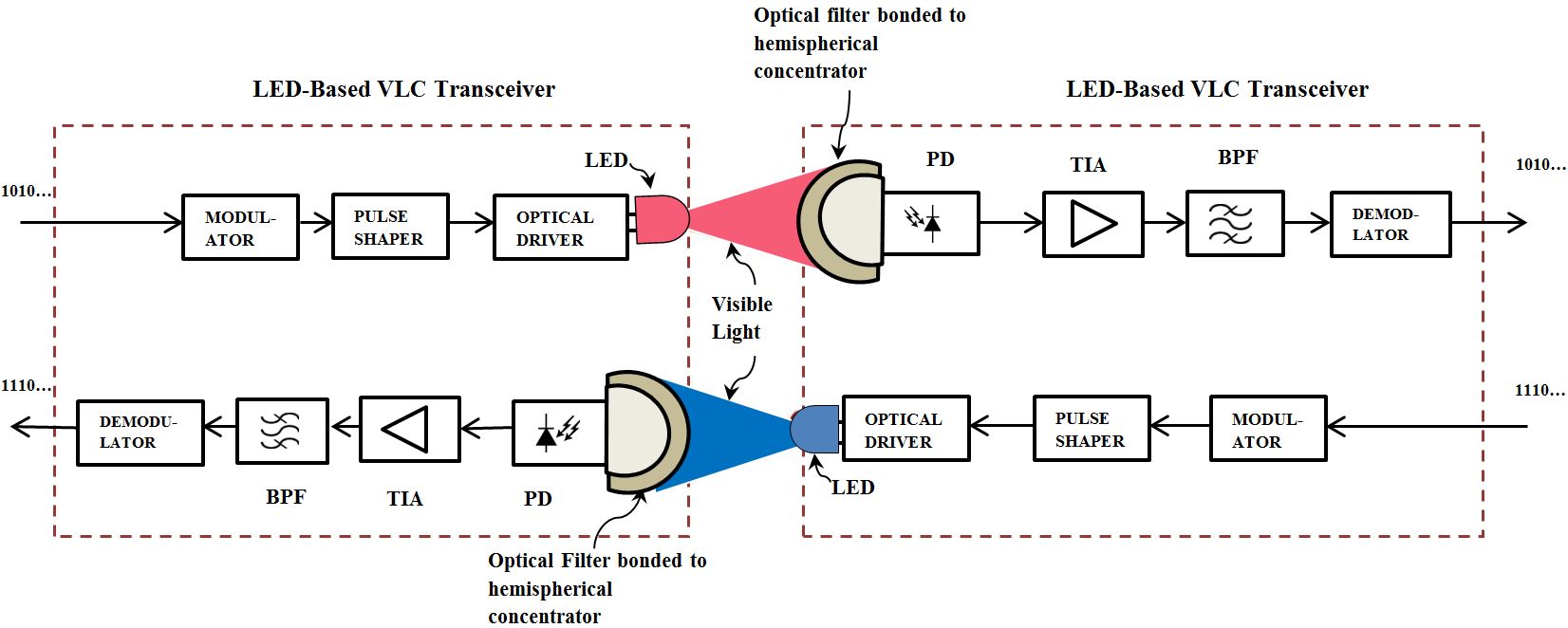}
  \caption{Conceptual Architecture of Full Duplex VLC System for ISC for Small Satellites}
\label{fig3}
\end{figure*}

\subsection{LED SPECIFICATION}


\begin{table}[!ht]
\caption{Spectral Colors Emitted By Specific Wavelengths} 
\centering 
\begin{tabular}{c c c c} 
\hline\hline 
Wavelength\ & Spectral Width\ & Bandwidth\ & Color\ \\
nm & nm & GHz \\[0.5ex] 
\hline 
\cellcolor{violet!15}381.5851 & \cellcolor{violet!15}0.1272 & \cellcolor{violet!15}262.1 & \cellcolor{violet!30} Violet\\ 
\cellcolor{violet!15}382.0436 & \cellcolor{violet!15}0.1712 & \cellcolor{violet!15}351.9 & \cellcolor{violet!30} Violet\\
\cellcolor{violet!15}382.5891 & \cellcolor{violet!15}0.1519 & \cellcolor{violet!15}311.3 & \cellcolor{violet!30} Violet\\
\cellcolor{violet!15}383.2310 & \cellcolor{violet!15}0.1685 & \cellcolor{violet!15}344.2 & \cellcolor{violet!30} Violet\\
\cellcolor{violet!15}383.8302 & \cellcolor{violet!15}0.1920 & \cellcolor{violet!15}391.0 & \cellcolor{violet!30} Violet\\
\cellcolor{violet!15}385.9922 & \cellcolor{violet!15}0.1554 & \cellcolor{violet!15}312.9 & \cellcolor{violet!30} Violet \\
\cellcolor{blue!25}393.3682 & \cellcolor{blue!25}2.0253 & \cellcolor{blue!25}3926.6 & \cellcolor{blue!55} Blue\\
\cellcolor{blue!25}396.8492 & \cellcolor{blue!25}1.5467 & \cellcolor{blue!25}2946.3 & \cellcolor{blue!55} Blue\\
\cellcolor{blue!25}410.1748 & \cellcolor{blue!25}0.3133 & \cellcolor{blue!25}558.7& \cellcolor{blue!55} Blue \\
\cellcolor{blue!25}434.0475 & \cellcolor{blue!25}0.2855 & \cellcolor{blue!25}454.6& \cellcolor{blue!55} Blue \\
\cellcolor{blue!25}486.1342 & \cellcolor{blue!25}0.3680 & \cellcolor{blue!25}467.2& \cellcolor{blue!55} Blue \\
\cellcolor{red!35}656.2808 & \cellcolor{red!35}0.4020 & \cellcolor{red!35}280.0& \cellcolor{red!55} Red\\ [1ex] 
\hline 
\end{tabular}
\label{table:table1} 
\end{table}

For our proposed system, LEDs with peak wavelengths centered in the blue and/ or red wavelengths can be utilized in the transmitter. Table II is an approximation of the spectral colors emitted by the wavelengths in Table I. Note that the boundaries depicted in the Table II are not precise.  The color of an LED is determined by the wavelength of the light emitted, which also depends on the semiconductor materials used in the manufacture of the LED. Thus, technically it is possible to manufacture LEDs for most wavelengths in the visible light band \cite{whitepaper}. The technology for creating Red and Green LEDs is generally viewed as mature. Aluminium gallium arsenide (AlGaAs) and gallium phosphide (GaP) can be used to manufacture red and green LEDs, respectively. With the development of aluminum indium gallium phosphide (AlInGaP), gallium nitride (GaN), and indium gallium nitride (InGaN), LEDs can be produced for a broad range of colors in the visible light spectrum. These new materials are now replacing GaP and AlGaAs as the semiconducting materials of choice for most commercial LEDs. These materials are durable and can withstand high temperatures which makes them ideal for space applications.

\subsection{PHOTODETECTORS}

Several factors influence the choice of a detector for a given application. Key among these include the light power level, wavelength range of the incident light, electrical bandwidth of the detector amplifier and the mechanical requirements of the application, such as size or temperature range of operation. Also important are cost, and the space environment. Most often, these criteria will limit the options for a given application. 

Avalanche photodiodes (APD) and PIN photodiodes have been used in many experimental studies on free space optical communication including VLC \cite{wood, barry, lee, kharraz}. APDs are advantageous over PIN photodiodes in applications where the dominant noise is the electrical noise in the pre-amplifier, rather than shot noise \cite{barry}. They have superior advantages in fiber optic systems, where the only source of shot noise is the photodetector dark current and the signal itself is weak. However, in free space optical communication systems, the background light is generally large enough that the resulting shot noise overshadows the thermal noise produced within the amplifiers and load resistors internal to the detection system (primarily in the front end), even with a PIN diode, thus limiting the usefulness of APDs for free-space optical wireless communication systems.

\section{PROPOSED SYSTEM DESCRIPTION}

Fig. 3 depicts a block diagram representation of the proposed LED-based VLC system for ISC.
The main sub-systems in the transmitter block are the modulator, optical driver and LED emitter; while the optical front-end, Si PIN photodetector (PD), transimpedance amplifier (TIA) and demodulator constitute the main elements in the receiver. The primary concept of the design is the utilization of Fraunhofer lines as natural low background noise channels for signal transmission. The design of the receiver optical front-end follows the approach proposed in \cite {barry} in order to take advantage of a high gain, wide FOV front-end. We provide further elaboration on the proposed transmitter and receiver front-end architectures.

\subsection{TRANSMITTER FRONT-END CONCEPT}

A single high-power LED or a bank of LEDs in series can be employed in VLC transmitter systems using On-Off Keying (OOK), which relies mainly on switching the light source on and off. However, OOK is a binary modulation scheme with low spectral efficiency. Thus, OOK can only provide limited data rates. Generally, optical transmitter front-ends using single high-power LEDs (or bank of LEDs) are not optimized for higher-order modulation and multi-carrier schemes. In \cite{haas}, the authors proposed an LED(s) transmitter front-end that is optimized for high data rates and can be used for higher-order modulation and multi-carrier schemes. They employed discrete power level stepping technique, which allows utilization of the full dynamic range of LEDs by avoiding non-linearity issues.

In this paper and for our simulations, we assume the transmitter front-end consist of a single LED or bank of low-power LEDs with an equivalent amount of ouput optical power.

\subsection{RECEIVER FRONT-END OPTICS}

The receiver front-end is designed as shown in Fig. 4.  A narrow-band optical filter is bonded to the outer surface of a hemispherical concentrator in order to achieve a high gain, wide FOV optical front-end. It was shown in \cite {barry} that, under certain conditions, the gain of the hemispherical front-end is nearly omni-directional which makes it a more useful configuration to deploy in a wide FOV application. It is also more robust to receiver movements and FOV misalignments compared to a planar optical front-end. We used PIN photodiode because the background light is generally large enough that the resulting shot noise dominates the thermal noise produced within the electrical front-end.

\begin{figure}[!ht]
\centering
\includegraphics[width=3.5in]{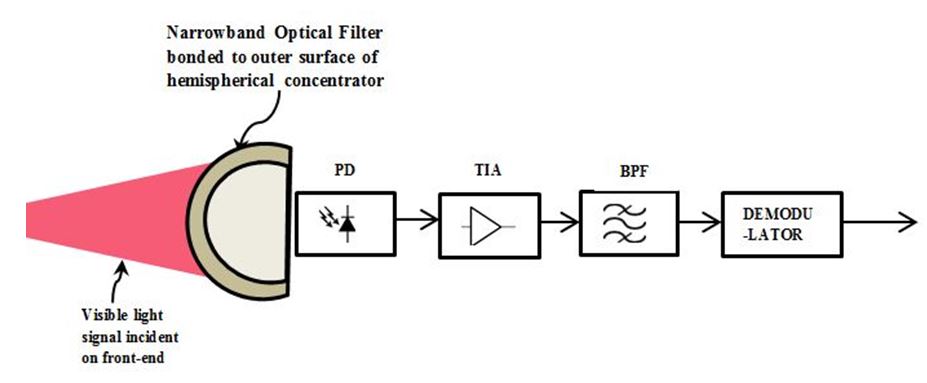}
\caption{Receiver Front-End Architecture}
\label{fig 4}
\end{figure}

\section{VLC LINK PHYSICAL MODEL}

We can model the line-of-sight (LOS) link between any two adjacent satellites in a trailing formation or within a cluster according to the generic LOS VLC scenario shown in Fig. 5. The distance between the LED emitter and detector is denoted by\begin{math}\ d \end{math}, while the detector aperture radius and physical area are represented by\begin{math}\ r \end{math} and\begin{math}\ A_{\pd} \end{math}, respectively. The angle of incidence with respect to the receiver axis is\begin{math}\ \psi \end{math}, and the angle of irradiance with respect to the transmitter perpendicular axis is\begin{math}\ \varphi \end{math}.  Angle\begin{math}\ \varphi \end{math} is referred to as viewing angle as it indicates how focused the beam is when emitted from the LED.

\begin{figure}[!ht]
\centering
\includegraphics[width=3.5in]{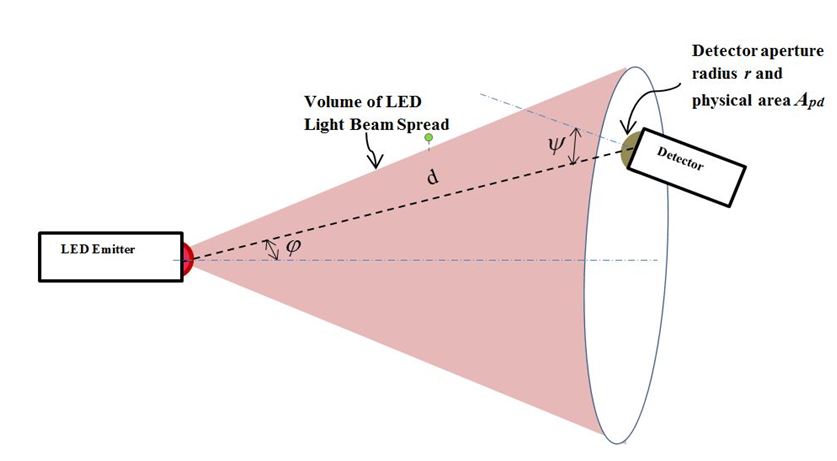}
\caption{LOS VLC Link Model: Adapted from \cite {cui}}
\label{fig 5}
\end{figure}

In line-of-sight (LOS) optical links, the relationship between the received optical power\begin{math}\ P_r \end{math} and the transmitted optical power\begin{math}\ P_t \end{math} can be represented by \cite{komine, amanor}

\begin{equation}\ P_r = H(0)P_t \end{equation}
The quantity\begin{math}\ H(0) \end{math}  represents the channel DC gain and it is the single most important quantity for characterizing LOS optical links.

As shown in \cite{barry2}, the channel gain in LOS optical links can be estimated fairly accurately by considering only the LOS propagation path and can be expressed as 

\begin{equation}
    H(0)=
    \begin{cases}
      \frac{(m+1)}{2\pi d^2} A_{\pd} \cos^m(\varphi) T_s g(\psi) \cos(\psi), &: 0\leq\psi\leq \psi_c\\
      0,&: \psi > \psi_c
    \end{cases}
 \end{equation}
where  \begin{math}\  m \end{math} is the order of Lambertian emission (i.e., a number which describes the shape of the radiation characteristics). The filter transmission coefficient (or gain) and concentrator gain are represented by the parameters \begin{math}\  T_s \end{math} and \begin{math}\  g(\psi) \end{math}, respectively, while the concentrator FOV semi-angle is denoted by \begin{math}\ \psi_c \end{math}. 

The Lambertian order \begin{math}\  m \end{math} is related to the semi-angle at half illuminance of an LED, \begin{math}\  \phi_{\frac{1}{2}} \end{math} and is given by \cite{cui}, \cite{barry2} 
\begin{equation}\  m  = \frac{-\ln2}{ \ln(\cos(\phi_{\frac{1}{2}}))} \end{equation}

By using a hemispherical lens (i.e., non-imaging concentrator) with internal refractive index\begin{math}\ n \end{math}, we can achieve a gain of \cite{barry}

 \begin{equation}
    g(\psi)=
    \begin{cases}
       \frac{n^2}{\sin^2 \psi_c} &: 0\leq\psi\leq \psi_c\\
      0,&: \psi > \psi_c
    \end{cases}
  \end{equation}
A hemisphere can achieve \begin{math}\  \psi_c \approx \frac{\pi}{2} \end{math}  and \begin{math}\  g(\psi) \approx n^2 \end{math} over its entire FOV provided the hemisphere is sufficiently large in relation to the detector, i.e., \begin{math}\  R > n^2r \end{math}, where \begin{math}\  r \end{math} and \begin{math}\  R \end{math} represents the detector and hemisphere radii, respectively \cite{barry2}. 

For a given receiver FOV, the effective signal-collection area\begin{math}\ A_{\eff}(\psi) \end{math} of the detector is given by\begin{math}\ A_{\eff}(\psi)  =  A_{\pd} \cos \psi  \end{math} where \begin{math}\ |\psi| < FOV \end{math}.

For non-Lambertian emission sources, (4) does not hold. For such sources, where the LEDs have particular beam shaping components, knowledge of the reshaped beam spatial distribution function \begin{math}\ g_s  (\theta) \end{math} is needed in order to calculate the path loss \cite{cui}.  

Following from (3), the average received optical power \begin{math}\  P_r \end{math} can be expressed as the sum of the transmitted power and path-loss on a dB scale, i.e., \begin{math}\ P_r  =  P_t + H(0)  \end{math}, 
where the channel has an optical path loss of \begin{math}\ -10\log_{10}H(0)  \end{math} [measured in Optical decibels]. 

The electrical signal component at the receiver side is given by \cite{lee}
\begin{equation}\ S = (\gamma P_r)^2  \end{equation}

Depending on the desired transmitter power, an array of standard LEDs can be used as the transmitter. When such multiple LEDs are used, spatially connecting distributed LEDs to a single receiver, we can obtain the total optical power by summing (or superimposing) the received power of all single LOS links within the receiver field of view (FOV) \cite{cui}. 

For the situation where two signals from different satellites are within the receiver's FOV, we can distinguish between these signals in the medium access control (MAC) layer. The MAC layer provides functionality for coordinating access to the shared wireless channel and utilizing protocols that facilitates the quality of communications over the medium. Interested readers are referred to \cite{radakrishnan2}, where various multiple access techniques applicable to ISC for small satellites systems are discussed.


\begin{table*}[!t]
\caption{Comparison of Different Models for Solar Flux Estimation} 
\centering 
\begin{tabular}{c c c c c c c c} 
\hline\hline 
No.\ &Wavelength Interval\ & Observed Solar Flux  \  & Solar Flux for a BB Sun \ & Proposed Model for Solar\  \\
-      &   (nm)                 & @ 1 AU (W/$\m^2$)     &  @ 5780K (W/$\m^2$)      & Flux @ 1 AU (W/$\m^2$)  \\[0.5ex] 
\hline 
1 &   240 - 400         &       118                                    & 158   &    157.18        \\ 
2 &   400 - 800         &       643                                    & 630   &    627.98         \\
3 &   800 - 1310       &       348                                    &  349  &    347.68          \\
4 &   1310 - 1860     &       148                                    & 123   &    122.92         \\
5 &   1860 - 2480     &        52                                     & 51    &    50.61           \\
6 &   2480 -3240      &       29                                      & 24    &   24.13        \\
7 &   3240 - 4500     &       17                                     & 14    &   13.95           \\
8 &   4500 - 8000     &   neglected                               &  7.7  &   7.70            \\
9 &   8000 - 12000    &  dust band                               & 1.3   &   1.30             \\
10 & 12000 - 24000  &  15 $\mu m $ $CO_2$ band        & 0.9    &   0.50             \\
11 & 24000 - 60000   &  neglected                              & 0     &    0.07                 \\
12 & 60000 - 1000000 & neglected                              & 0      &   0.00           \\

\hline 
\end{tabular}
\label{table:table2} 
\end{table*}


\section{The Noise Model}

In this work, we consider the Sun as the main source of background illumination from the environment. We modeled the Sun as a blackbody using Planck's blackbody radiation model, in which spectral irradiance of the source is a function of wavelength and temperature \cite{lee}, i.e.,
\begin{equation}\  W(\lambda,T) = \frac{2\pi h_p c^2}{ \lambda^5} \frac{1}{(e^{\frac{h_pc}{\lambda k T}}-1)}   \end{equation}
where \begin{math}\  \lambda \end{math} is the wavelength, \begin{math}\  c \end{math} is the speed of light, \begin{math}\  h_p \end{math} is Planck\textquotesingle s constant, \begin{math}\  k \end{math} is Boltzmann’s constant and \begin{math}\  T \end{math} is average temperature of the Sun\textquotesingle s surface. 

Following the approach of Spencer \cite{spencer},  we developed a simple yet fairly accurate analytical model that describes the irradiance that falls within the spectral range of the receiver optical filter

\begin{equation}\ E_{det} \approx  2.15039 \times 10^{-5} d_f t_f  \int_{\lambda_a}^{\lambda_b} W(\lambda,T) d\lambda  \end{equation} 
where  \begin{math}\  d_f \end{math}   and  \begin{math}\  t_f \end{math} are coefficients that represents the \emph{day of the year} and \emph{time of day}, respectively. For this work, we assume the maximum value for\begin{math}\  t_f \end{math}, which is 1.0.

We validated our model by evaluating (9) for different wavelength intervals and compared the results with observed solar fluxes  (W/$\m^2$)  taken from the 1985 Wehrli Standard Extraterrestrial Solar Irradiance Spectrum and a Blackbody (BB) Sun model from NASA \cite{nasa},\cite{nasamodel}. The BB Sun produces an integrated flux over these intervals of 1359 W/$\m^2$ at 1 astronomical unit (AU) compared to 1355 W/$\m^2$ for the observed sun. Our model produces an integrated flux of 1354 W/$\m^2$ over the same wavelength intervals as shown in Table III.

The background noise power detected by the optical receiver physical area can be computed as \cite{barry}:

\begin{equation}\ P_{bg} =E_{det} T_s A_{\pd} n^2   \end{equation} 
where \begin{math}\  T_s \end{math} is the filter transmission coefficient and\begin{math}\  n \end{math} is the internal refractive index of the concentrator at the receiver\textquotesingle s optical front-end. 

The total input noise variance \begin{math}\  N \end{math} is the sum of the variances of the shot noise and thermal noise \cite{barry}:
\begin{equation}\ N = \sigma^2_{\shot}  + \sigma^2_{\thermal}  \end{equation} 

We neglect the effects of intersymbol interference (ISI) based on the assumption that the inter-satellite link between any two adjacent satellites in a leader-follower or cluster formation is not susceptible to multipath propagation. 

The shot noise variance is given by \cite{lee}
\begin{equation}\ \sigma^2_{\shot} = 2q\gamma(P_r + I_2 P_{bg})B  \end{equation} 
where \begin{math}\  q \end{math} is the electronic charge, \begin{math}\  B \end{math} is the equivalent noise bandwidth, \begin{math}\  \gamma \end{math} represents the photodetector responsivity, and \begin{math}\  I_2 \end{math} is the noise bandwidth factor for a rectangular transmitter pulse. 

Following the analysis in \cite{barry}, the thermal noise variance can be expressed by:
\begin{equation}\ \sigma^2_{\thermal} = \frac{8\pi k T_{\A}}{G} \eta A_{\pd} I_2 B^2  +  \frac{16\pi^2 k T_{\A} \Gamma}{g_m} \eta^2 A^2_{\pd} I_3 B^3 \end{equation} 
where \begin{math}\  k \end{math} is Boltzmann’s constant, \begin{math}\  T_{\A} \end{math} is the absolute temperature, \begin{math}\  G \end{math} is the open-loop voltage gain,\begin{math}\  \eta \end{math} is the fixed capacitance of photodetector per unit area, \begin{math}\  \Gamma \end{math} is the FET channel noise factor, \begin{math}\  g_m \end{math} is the FET transconductance and \begin{math}\  I_3 \end{math} is the noise bandwidth factor for a full raised-cosine pulse shape \cite{barry}.

Finally, the electrical SNR at the receiver, which is a key metric for measuring the quality of the communication link, can be determined by

\begin{equation}\ \SNR =\frac{S}{N}= \frac{(\gamma P_r)^2}{\sigma^2_{\shot}  + \sigma^2_{\thermal}}   \end{equation}

\section{CHARACTERISTICS OF THE VLC MODULATED SIGNAL}

A key difference between VLC and RF communications is in the way data is encoded or conveyed. While data can be encoded in the amplitude or phase of an RF signal, signal intensity is the primary parameter used for conveying information in VLC systems \cite{pathak, tsonev}. The implication is that phase and amplitude modulation techniques cannot be applied in VLC; rather the data has to be encoded in the varying intensity of the emitting light pulses \cite{tsonev}. At the receiver side, direct detection is the dominant approach for signal recovery due to changes in the instantaneous power of the transmitted signal \cite{medina}. Thus,  IM/DD schemes are the main modulation/demodulation methods used in VLC systems. A further attribute of an IM/DD system is that the modulating signal must be both {\em real valued} and {\em unipolar} \cite{tsonev}. This distinctive feature of VLC, as an IM/DD system, has profound consequence on the type of modulation scheme to use. In other words, many full-fledged modulation schemes used in RF communications are inapplicable in VLC systems. Additionally, unlike RF communication systems, the modulation scheme for a VLC system is generally required to support dimming and flicker mitigation \cite{pathak}. Dimming is particularly important for applications where illumination is not a primary requirement as it can be used as a technique for conserving energy and increasing battery life. Nevertheless, dimming should not result in degradation of the communication performance. Besides dimming, an additional requirement for VLC modulation schemes is resistance to flickering. Flickering is the human-perceivable fluctuations in the brightness of light and it is usually caused by long runs of 0s or 1s in the data sequence which can reduce the rate at which light intensity changes and cause the flickering effect \cite{pathak}. Flickering was shown in \cite{berman} as a likely cause of adverse physiological changes in humans. However, for a space-based application, flickering may not be an issue.

\section{LINK BUDGET DESIGN}

Unlike RF communication links, not much work has been done in the formulation and analysis of link budgets for visible light links between CubeSats. The closest work in the literature is the seminal work done by \cite{popescu}, where they examined the power budgets for inter-satellite links between CubeSat radios. However, the link budget parameters for an RF link differ from a VLC link. While the propagation path loss of an RF link is dependent on the radio signal frequency, path loss for LOS optical links is assumed to be independent of wavelength. 

Following from (3), (7), (11) and (14), the SNR per bit can be expressed as \cite{ghassemlooy},\cite{sklar}:

\begin{equation}\ \SNR =\frac{E_b}{N_o}= \frac{[\gamma H(0) P_t ]^2}{N} \frac{B}{R}  \end{equation} 
where \begin{math}\   B \end{math} is the bandwidth in Hz over which noise is measured,  \begin{math}\  R \end{math} represents the desired bit-rate to be supported by the link in bits per second (bps), and \begin{math} \frac{E_b}{N_o}  \end{math} is the bit-energy per noise-spectral-density. Note also that \begin{math} N = N_o B  \end{math} \cite{sklar}, where \begin{math} N_o  \end{math}, is the maximum single-sided noise power spectral density in W/$\Hz $, and it is generally assumed to be uniformed.  

Equation (15) can be expressed on a logarithmic dB scale, which is a more appropriate form for the analysis of the link power budget

\begin{equation}\ \SNR (\dB) = 10 \log_{10} \left( \frac{[\gamma H(0) P_t ]^2}{N} \frac{B}{R} \right) \end{equation} 

    \begin{flalign}
        \SNR (\dB) &= 10 \log_{10} \gamma^2 + 10 \log_{10} H(0)^2  + 10 \log_{10} (P_t)^2  + && \\\nonumber
         & 10 \log_{10} B -10 \log_{10} N - 10 \log_{10}R  &&
   \end{flalign}

From (17), it is possible to estimate the minimum transmitter power required to achieve  a targeted SNR. To ensure a resilient link, the link budget usually include other terms to account for additional losses as well as a link margin.

\section{PERFORMANCE EVALUATION AND RESULTS}

For our system model, we consider two, 1U CubeSats in direct LOS and in a leader-follower configuration. We assume that the satellites are deployed in nearly circular lower Earth orbits and that the distance between the CubeSats is fixed. We used the numerical values in Table IV for the simulation of our analytical model. The optical filter at the receiver\textquotesingle s front-end is tuned to the deep Fraunhofer line at 656.2808 nm wavelength with a line width of 0.4020 nm. We assumed a concentrator radius of 2.0 cm and PIN photodiode with active physical area of 7.84 $\cm^2$  (Hamamatsu Si Photodiode S3584). 

\begin{table}[ht]
\caption{Simulation Model Parameter Assumptions} 
\centering 
\begin{tabular}{l c} 
\hline\hline 
Parameter\ & Value\ \\
\hline 
Semi-angle at Half Power, \begin{math}\  \Phi_{\frac{1}{2}} \end{math}                          &      \begin{math}\  30^o \end{math}  \\
LED Peak Wavelength, \begin{math}\  \lambda_{\peak} \end{math}                        &         656.2808 nm \\
Concentrator FoV Semi-angle, \begin{math}\  \psi_c \end{math}                         &        \begin{math}\ 35^o \end{math}\\
Filter Transmission Coefficient, \begin{math}\ T_o \end{math}                         &         1.0 \\
Incidence Angle, \begin{math}\  \varphi \end{math}                         &         \begin{math}\  30^o \end{math} \\
Irradiance Angle, \begin{math}\  \psi \end{math}                          &         \begin{math}\  15^o\end{math} \\
Detector Responsivity, \begin{math}\ \gamma \end{math}                         &        \begin{math}\  0.51 \end{math}\\
Refractive Index of Lens, \begin{math}\ n \end{math}                      &         1.5  \\
Radius of Concentrator, \begin{math}\ R \end{math}                       &         2.0 cm \\
Detector Active Area, \begin{math}\ A_{\pd} \end{math}                      &       7.84\begin{math}\ \cm^2 \end{math} \\

Desired Electrical Bandwidth, \begin{math}\  B \end{math}                          &         0.5 MHz \\
Optical Filter Bandwidth, \begin{math}\ \Delta \lambda \end{math}                         &     0.4020 nm  \\
Optical Filter Lower Limit, \begin{math}\ \lambda_1 \end{math}                         &     656.0798 nm  \\
Optical Filter Upper Limit, \begin{math}\ \lambda_2 \end{math}                         &     656.4818 nm  \\

Open Loop Voltage Gain, \begin{math}\ G \end{math}                       &         10 \\
FET Transconductance, \begin{math}\ g_m \end{math}                      &         30 ms \\
FET Channel Noise Factor, \begin{math}\ \Gamma \end{math}                       &         0.82 or 1.5  \\
Capacitance of Photodetector, \begin{math}\ \eta \end{math}                      &        38 pF\begin{math}\ /\cm^2 \end{math}  \\
Link Distance, \begin{math}\ d \end{math}                       &         0.5 km \\

Noise Bandwidth Factor for White Noise,\begin{math}\ I_2 \end{math}                       &         0.562 \\
Noise Bandwidth Factor for \begin{math}\ f^2 \end{math} noise, \begin{math}\  I_3 \end{math}                     &         0.0868 \\
Boltzmann Constant, \begin{math}\ k \end{math}                       &   \begin{math}\ 1.3806\times10^{-23}\J/\K\end{math}\\
Absolute Temperature, \begin{math}\ T_{\A} \end{math}                      &        300 K \\

\hline\hline 
\end{tabular}
\label{table:table1} 
\end{table}


In this section, we investigated the impact of solar background illumination on the SNR at the receiver, and then conducted a comparative evaluation of the ISC link performance for five different IM/DD schemes, namely, on-off keying non-return-to-zero (OOK-NRZ), pulse position modulation (PPM), digital pulse interval modulation (DPIM), DC biased optical OFDM (DCO-OFDM) and asymmetrically clipped optical OFDM (ACO-OFDM). These schemes were considered based on the individual merits they bring to small satellites. These include bandwidth and power efficiency, reduced implementation complexity, as well as robustness to ISI. We also assessed the performance of the VLC link with and without the use of forward error correction (FEC). 

Table V is a summary of methods for determining the BER and bandwidth requirements for the above modulation schemes. The BER has been expressed as a function of SNR to simplify the analysis and allow a quantitative comparison of the different schemes. 


\begin{table}[!ht]
\caption{Methods for BER and Bandwidth Requirements \cite {trisno,mesleh,elganimi,hayes,amanorphd}} 
\centering 
\begin{tabular}{c c c } 
\hline\hline 
Modulation\ & BER\ & Bandwidth \  \\
Scheme       &    & Requirement   \\[0.5ex] 
\hline 
\\[0.1ex]
OOK-NRZ      &  \begin{math}\ \frac{1}{2} \erfc(\frac{1}{2\sqrt{2}} \sqrt{\SNR})    \end{math}                                          &  \begin{math}\  R_b \end{math} \\ 
\\[0.1ex]
L-PPM           &  \begin{math}\ \frac{1}{2} \erfc(\frac{1}{2\sqrt{2}} \sqrt{\SNR \frac{L}{2} \log_2 L })\end{math}                 &   \begin{math}\  R_b \frac{L}{\log_2 L} \end{math} \\
\\[0.1ex]
DPIM          &  \begin{math}\ \frac{1}{2} \erfc(\frac{1}{2\sqrt{2}} \sqrt{\SNR \frac{L_{avg}}{2} \log_2 L })\end{math}         &   \begin{math}\  R_b \frac{L_{avg}}{\log_2 L} \end{math} \\
\\[0.1ex]
DCO-OFDM   &  \begin{math}\ \frac{\sqrt{M} -1}{\sqrt{M} \log_2 \sqrt{M}} \erfc(\sqrt{\frac{3 \SNR}{2(M-1)} }) \end{math}  &  \begin{math}\  \frac{R_b(N+N_g)}{(\frac{N}{2}-1)\log_2M} \end{math}    \\
\\[0.1ex]
ACO-OFDM   &    \begin{math}\ \frac{\sqrt{M} -1}{\sqrt{M} \log_2 \sqrt{M}} \erfc(\sqrt{\frac{3 \SNR}{2(M-1)} }) \end{math}  &  \begin{math}\  \frac{R_b(N+N_g)}{(\frac{N}{4}-1)\log_2M} \end{math}      \\

\\[0.1ex]
\hline\hline 
\end{tabular}
\label{table:table1} 
\end{table}

\subsection{Impact of Solar Background on SNR}

For a transmitted optical power of 2W, Fig. 6 represents a plot of the SNR for different values of concentrator FoV. Clearly, the impact of the concentrator FoV on the SNR is apparent. A \begin{math}\ 10^o \end{math} reduction in the FoV semi-angle translates into an improvement of the SNR by about 3.5 dB. It is important, following the analysis in \cite{barry}, that the concentrator FoV semi-angle,\begin{math}\ \psi_c \end{math} is greater than the incidence angle,\begin{math}\ \varphi \end{math} in order to achieve a concentrator gain \begin{math}\ g(\psi) \end{math} of \begin{math}\ n^2 \end{math} or greater. Fig. 7 shows that doubling the link distance results in a drastic degradation of SNR. It is also evident from (3), (4) and (14), that doubling the transmitted optical power or halfing the active detector area has a profound impact on SNR. However, for a given small satellite configuration, the SMaP constraints limit the extent to which \emph{power} and \emph{detector area} can be extended. Using the minimum desired bandwidth for a given application will also yield an improved SNR. Ultimately, the task of the communication system designer is to trade-off these critical parameters in order to achieve the desired performance.

\begin{figure}[!t]
\centering
\includegraphics[width=3.3in, height=5.6cm]{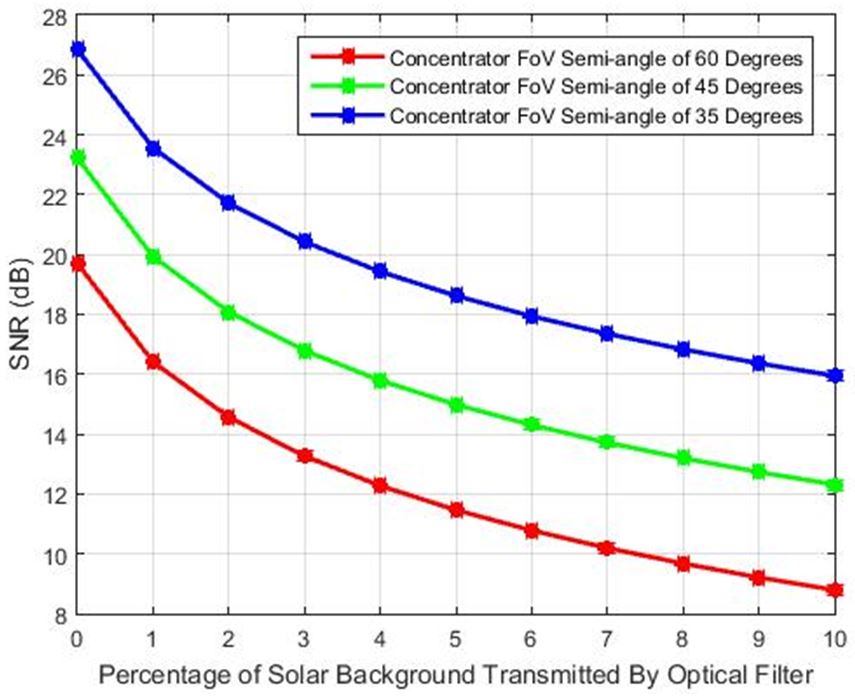}
\caption{Impact of Solar Background on SNR for Link Distance of 0.5 km, Transmitted Optical Power Output of 2W and Electrical Bandwidth of 0.5 MHz}
\label{fig 6}
\end{figure}

\begin{figure}[!t]
\centering
\includegraphics[width=3.3in, height=5.6cm]{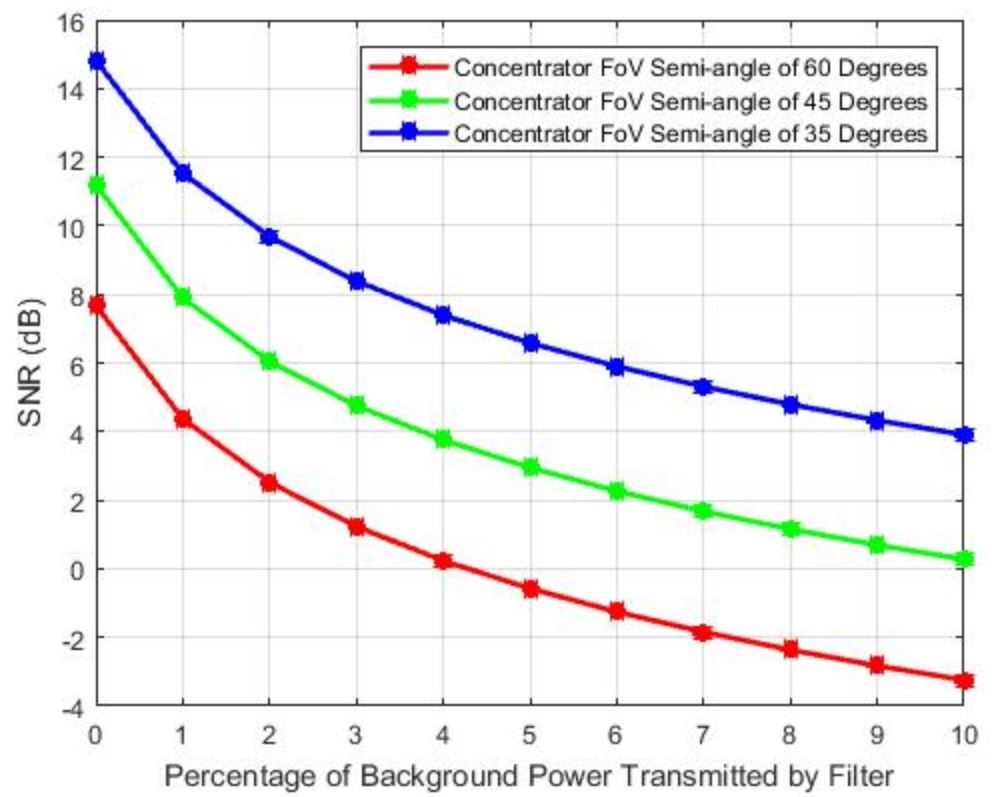}
\caption{SNR Plot for Link Distance of 1.0 km, Transmitted Optical Power Output of 2W, and Electrical Bandwidth of 0.5 MHz for Different Concentrator FoVs}
\label{fig 7}
\end{figure}

\subsection{Analysis of Different IM/DD Schemes}

For a targeted BER of 10\textsuperscript{-6}, Table VI depicts the required transmitted optical power for the different IM/DD modulation schemes. The results show that for higher levels of L (i.e., L \begin{math} \geq 4  \end{math}), PPM requires less optical power than OOK-NRZ to achieve the same error performance. Similarly, for L=8, DPIM requires 65 percent less optical power than OOK. Moreover, unlike PPM, DPIM requires no symbol synchronization, thus yielding a less complicated receiver structure. Compared to multi-carrier modulation schemes such as DCO-OFDM and ACO-OFDM, DPIM (L=8) requires about 67 percent less power than ACO-OFDM (M=16) for the same BER. As illustrated in Fig. 8, at low to moderate data-rates, PPM and DPIM exhibit better error properties than DCO-OFDM and ACO-OFDM. However, at very high data-rates, the multi-carrier schemes (i.e., DCO-OFDM and ACO-OFDM) are more resilient to noise and offer superior capabilities in terms of throughput. 

\begin{figure}[!ht]
\centering
\includegraphics[width=3.4in, height=5.6cm]{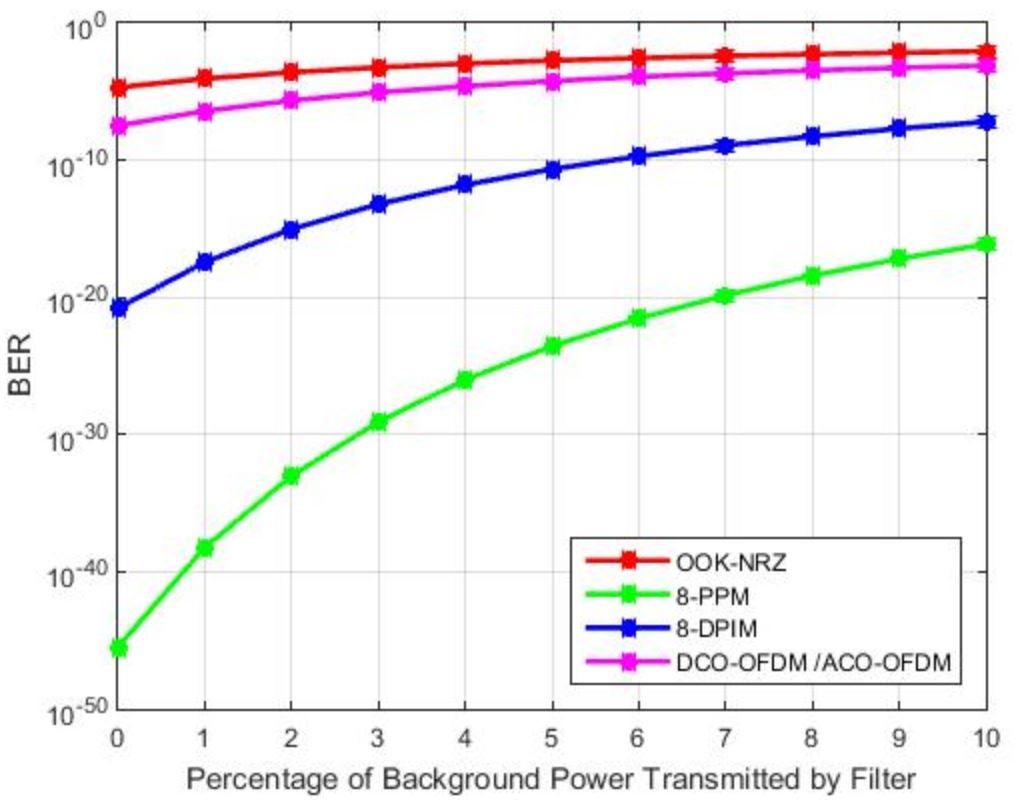}
\caption{BER Plot for Link Distance of 0.5 km, Transmitted Optical Power Output of 4W, and Electrical Bandwidth of 2.5 MHz}
\label{fig 8}
\end{figure}

The disadvantage of these schemes is the cost of  the associated high transmitted optical power. Clearly, for low to moderate data-rates, the higher power requirement of DCO-OFDM, puts it at a relative disadvantage to power-efficient modulation schemes required for small satellites, where mass and volume of onboard electronics are restricted. The simplified receiver structure of DPIM coupled with its relatively good power-efficiency and bandwidth requirements makes it an attractive choice for ISC for small satellites at moderate data-rates. For very high data-rates, the multi-carrier schemes can be considered at the expense of high transmitted optical power.


\begin{table}[!t]
\caption{Required Transmitted Optical Power for Link Distance of 0.5 km, Assumed Bandwidth of 0.5 MHz and Targerted BER of 10\textsuperscript{-6}} 
\centering 
\begin{tabular}{c c c c  } 
\hline\hline 
Modulation \ &  \ & SNR\ & TX Optical Power \ \\
 Scheme  &   & (dB) & @ 5 Percent Background  \\[0.5ex] 
\hline 

OOK-NRZ    &                &  19.56   & 2.2 W    \\ 
\hline 

L-PPM        &   L=2       & 19.56    & 2.2 W     \\
                &   L=4       & 13.54    & 1.1 W    \\
                &   L=8       & 8.77     & 0.6 W      \\
\hline 

DPIM        &   L=2       & 18.59    & 1.97 W    \\
                &   L=4       & 14.12    & 1.18 W    \\
                &   L=8       & 10.40    & 0.77 W    \\
\hline 

DCO-OFDM   &   M=4       & 13.54    & 1.1 W + DC Bias   \\
                   &   M=16     & 20.42    & 2.4 W + DC Bias    \\
                   &   M=64     & 26.56    & 5.0 W + DC Bias     \\
\hline 

ACO-OFDM   &   M=4       & 13.54    & 	1.1 W    		\\
                   &   M=16     & 20.42    & 	2.4 W    		\\
                   &   M=64     & 26.56    & 	5.0 W    		\\

\hline\hline 
\end{tabular}
\label{table:table2} 
\end{table}

\subsection{Uncoded versus Coded Performance Evaluation}

In this sub-section, we examined the impact of forward error-correction (FEC) on the performance of the VLC link. We used the uncoded transmission characteristic of a 16-QAM constellation, which can be applied in ACO-OFDM and DCO-OFDM schemes. The simulation was carried out in MATLAB for a range of bit-energy per noise-spectral-density Eb/No (i.e., SNR per bit) values from 7dB to 12 dB.

For the coded case, we used a Reed-Solomon encoder and decoder pair consisting of a RS(15,11) code. The code has two-symbol error correction capability and a generator polynomial given by:


\begin{equation}
\label{eq:1}
\begin{aligned}
g(X)= X^4 + (\alpha^3 + \alpha^2 + 1)X^3 + (\alpha^3 + \alpha^2)X^2  \\
+ (\alpha^3)X + (\alpha^3)X + (\alpha^2 + \alpha + 1),
\end{aligned}
\end{equation}

where \begin{math} \alpha \end{math} is root of the primitive polynomial \begin{math} p(X) \end{math} in GF(16):

\begin{equation} p(X)= X^4 + X + 1 \end{equation}

Thus, the generator polynomial $ g(X) $ can be expressed as:

\begin{equation} g(X)= X^4 + 13X^3 + 12X^2 + 8X + 7  \end{equation}

We further examined the impact of redundancy on the BER by comparing the performance of a RS(15,13) encoder/decoder pair against the above encoder/decoder pair and the uncoded modulation case. The generator polynomial of the RS(15,13) code is given by 

\begin{equation} g_2(X)= X^2 + (\alpha^2 + \alpha)X + \alpha^3   \end{equation}

where \begin{math} \alpha \end{math} is root of primitive polynomial (24) in GF(16), \\
i.e., \begin{equation} g_2(X)= X^2 + 6X + 8  \end{equation}


\begin{figure}[!ht]
\centering
\includegraphics[width=3.4in, height=5.6cm]{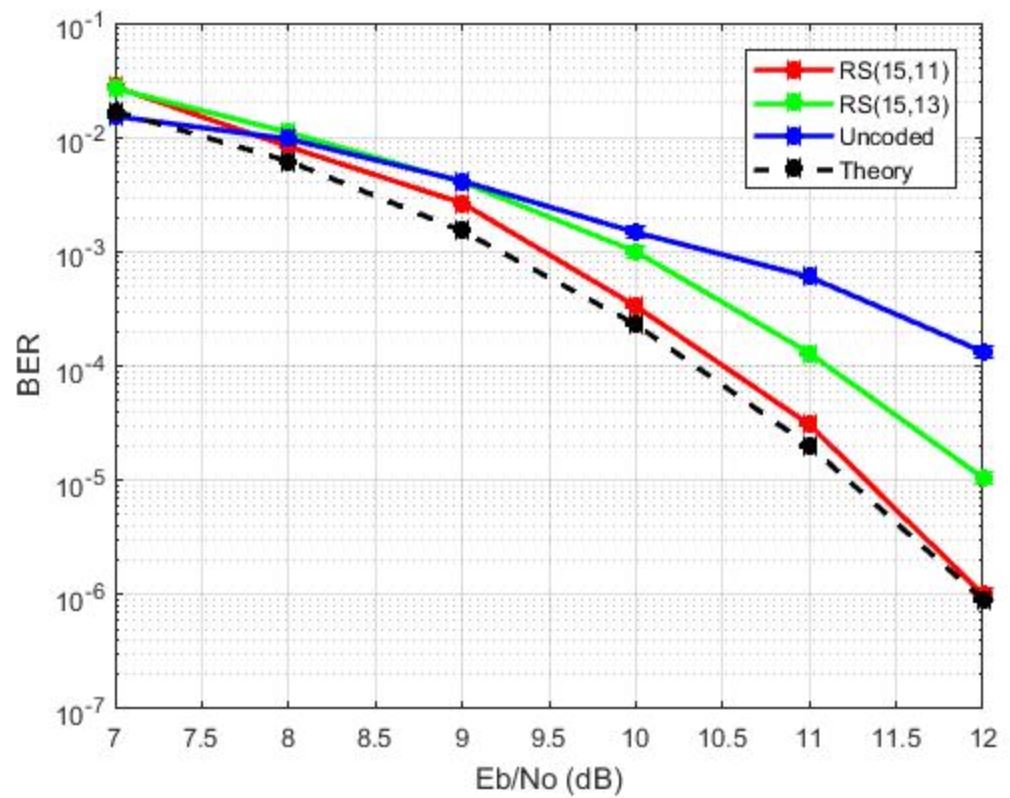}
\caption{Bit Error Rate versus Eb/No (i.e., SNR per bit) }
\label{fig 9}
\end{figure}

Fig. 9 depicts the simulation results of the uncoded and coded cases. For a Eb/No of 12 dB, the error probability of RS (15, 11) has improved by a factor of more than 100 compared to the uncoded case. Clearly, the added redundancy resulted in faster signaling, less energy per channel symbol, and more errors detected out of the demodulator. It is evident from the profile of the RS (15, 13) that the higher the redundancy (i.e., the lower the code rate), the better the bit-error performance. However, the implementation complexity of a RS encoder rises with increases in redundancy. Additionally, there must be corresponding expansion in bandwidth to accommodate the redundant bits for any real-time communications application.

\section{CONCLUSIONS}

A major limitation of small satellites is their restricted form factor which regulates the size, mass, and power of the electronics that can be carried onboard. For a given small satellite configuration, these restrictions limit the range and throughput that can actually be achieved across the ISL. 

In this paper, we proposed an LED-based VLC system for ISC that addresses the SMaP constraints of small satellites and discussed essential physical layer requirements and design concepts for the realization of high performance visible light ISLs. The proposed system can be deployed within a constellation of small satellites and it is capable of establishing reliable communication links in the presence of steady background solar radiation through the use of natural low-background noise channels.

The major contributions of this work include the following:

\begin{enumerate}

\item This work is the first to provide a quantitative assessment of solar background illumination on ISLs between small satellites.

\item We investigated the use of natural low background noise channels (i.e., Fraunhofer lines) for VLC systems of medium scope using hypothetical LEDs whose peak wavelength coincides with the chosen Fraunhofer lines. 

\item We developed an analytical model of the ISL and evaluated the impact of solar background illumination on its performance for both uncoded and coded IM/DD schemes.

\item The work discussed the design and formulation of power link budget for VLC ISLs.

\item We discussed physical layer design issues and attempt to provide recommendations on key issues to be considered in the development of VLC-based communication subsystem for multiple small satellite systems. 

\end{enumerate} 

Using a transmitted optical power of 4W  and DPIM modulation, a receiver bandwidth requirement of 3.5 MHz is needed to achieve a data rate of 2.0 Mbits/s for a moderate link distance of 0.5 km at an uncoded BER of 10\textsuperscript{-6}, which is the performance requirement for stable communication link. This data rate is sufficient to support navigation, command and health data as well as science data.


%





\ifCLASSOPTIONcaptionsoff
  \newpage
\fi




%
%

\bibliographystyle{IEEEtran}
\bibliography{IEEEabrv,journal_paper}



%

%
%
%
%
%
%
%





\end{document}